\documentclass[final,3p,12pt]{elsarticle}

\usepackage{graphicx}

\usepackage{amssymb}

\usepackage{lineno}

\usepackage{verbatim}
\usepackage{subcaption}
\usepackage{url}

\journal{}

\begin{document}

\begin{frontmatter}



\title{A Century of Light-Bending Measurements: Bringing Solar Eclipses into the Classroom}


\author[anv]{Emanuele Goldoni\corref{corr}}
\author[anv]{Ledo Stefanini}
\cortext[corr]{Corresponding author: \textbf{\texttt{emanuele.goldoni@gmail.com}}}
\address[anv]{Accademia Nazionale Virgiliana, 46100 Mantova, Italy}

\begin{abstract}
In 1919, Eddington and Dyson led two famous expeditions to measure the bending of light during a total solar eclipse. The results of this effort led to the first experimental confirmation of Einstein's General Relativity and contributed to create its unique and enduring fame. Since then, similar experiments have been carried out all around the world, confirming the predictions of the General Relativity. Later, developments in radio interferometry provided a more accurate way to measure the gravitation deflection. We believe that - after more than a century - starlight deflection caused by the Sun's gravity still represents a simple and intuitive way to introduce high-school students to General Relativity and its effects.
To this aim, we gathered measurements taken during eight eclipses spanning from 1919 to 2017, and we created a single dataset of homogeneous values. Together with the whole dataset, this article provides a blueprint for a possible group activity for students, useful to introduce the theory in physics classes with a playful approach.
\end{abstract}


\end{frontmatter}

\section{Introduction}\label{sec:introduction}

The first decades of the last century saw the emergence of two of the greatest scientific revolutions: the Theory of Relativity and Quantum Mechanics. The former was the exclusive work of Albert Einstein, while the latter was the joint work of Erwin Schr\"{o}dinger and Werner Heisenberg. These two fundamental theories for our life were born almost simultaneously, but had very different roots. Quantum Mechanics was meant to give rational structure to a great collection of experimental facts that did not fit in the classical theories of mechanics and electromagnetism. On the other hand, General Relativity came from a deep need to collect the space-time stage of physical phenomena in an aesthetic synthesis. Einstein worked on this huge project for about ten years starting from 1906, passing through great hopes and atrocious disappointments, before arriving at a mathematically simple and aesthetically dazzling synthesis of its theory. 

In 1911 Einstein himself suggested the idea that light could be influenced (and bent) by a gravitational field: he calculated that the observed position of a star whose light passed near the Sun's limb would change by $0.87 \textrm{ arcsec}$.
Since 1912, a few expeditions were scheduled to confirm (or disprove) General Relativity. Luckily for the famous physicist, bad weather and war prevented the astronomers from making any observations \cite{Coles2001}. In 1915, in fact, Einstein found that the general Principle of Equivalence needed a modification of the Newtonian law of gravitation and adjusted his predictions: the correct light displacement value was not the same as the Newtonian result, but twice as large ($1.75 \textrm{ arcsec}$).

At the end of the Great War, a new opportunity for an experimental verification of General Relativity arose: a solar eclipse on May 29, 1919 was going to happen just when the Sun would have been in front of unusual bright stars belonging to the Hyades cluster. The English Royal Society, in the person of the astronomer Arthur Eddington, seized the opportunity and arranged two expeditions to take accurate pictures \cite{Dyson1920}.

The results, confirming Einstein's generalized relativity theory, were presented by Eddington to the Royal Society of London and made the front page of many worldwide newspapers. Headlines like ``Revolution in science: new theory of the universe: Newtonian ideas overthrown'' (Times) and ``Lights all askew in the heavens: Einstein's theory triumphs'' (The New York Times) transformed the father of General Relativity into a global celebrity.

Although the 1919 experiment had an immediate impact on Einstein's popularity, the acceptance of his theory by the worldwide scientific community was slower: the high uncertainty in Eddington's measurements and a poor understanding of general relativity among other scientists at the time fuelled a controversy that surrounded for years the outcome of the expeditions.

During the 1920s, scepticism about General Relativity theory and its predictions continued until new observations provided much more convincing statistical data, settling (relatively) the matter. Optical measurements of light deflection during eclipses continued until the 70s, when radio interferometry provided a more accurate way to measure the gravitation deflection.

No more expeditions have been organized after 1973, but in 2017 the amateur astronomer Donald Bruns replicated the eclipse experiment using commercially available equipment and obtaining a result in perfect agreement with Einstein's predictions. As Burn commented on his achievement, ``\emph{While there is no new science resulting from this experiment, the hopes of the 20th century astronomers have been realized}'' \cite{Bruns2018}.

In November 2019, a series of public lectures organized by the Accademia Nazionale Virgiliana and the local chapter of AIF (Associazione per l'Insegnamento della Fisica) to celebrate the anniversary of the famous Dyson's announcement prompted us to gather all the data taken by astronomers during the years.
A careful bibliographic research led us to collect different sets of measurements, but we had not been able to find a single dataset containing all the values. In 1960 H. Von Kl\"{u}ber summarized the results obtained by eclipse expeditions until 1959 and represented graphically the measures \cite{Kluber1960}. Unfortunately, his article only includes low-resolution plots and does not to provide the numerical values. Hence, we transcribed all the measurements from the original papers, we harmonized the values and we created a single dataset of homogeneous observations. Finally, we decided to release it publicly, believing it could be useful for teachers.

In this article we briefly describe the expeditions which provided the measurements and we suggest a possible playful activity for students based on our dataset. We hope that playing with the same data that Eddington, Dyson, Freundlich and Van Biesbroeck handled in the past could make physics appear less scaring to students and could inspire the next generation of scientists.

\section{The Expeditions}\label{sec:expeditions}

Our database includes the measurements collected by astronomers during total eclipses in 1919, 1922, 1929, 1947, 1952, 1973, and 2017.
Three notable expedition are still missing in our project. The results obtained during the eclipse of 21 September 1922 by Campbell and Trumpler, \cite{Campbell1923} and by Dodwell and Davidson \cite{Dodwell1924} are currently under analysis and will be integrated in the future. In addition, we have not been able to find the original article by A. Mikhailov on the eclipse in USSR in June 1936. 

A brief description of the eight considered expeditions will follow: for more details on the technical equipment used, the methodology applied and the anecdotes of these scientific missions we suggest to refer to the original articles pointed in the references.

We are aware that, quite often, the methodologies used in the original works were later disputed by other scientists and new values were derived from the same observation \cite{Kluber1960}. In all cases, we preferred to stay with the original measurements.

\begin{figure}[htb]
\centering
\begin{subfigure}[h]{0.45\linewidth}
\includegraphics[width=\linewidth]{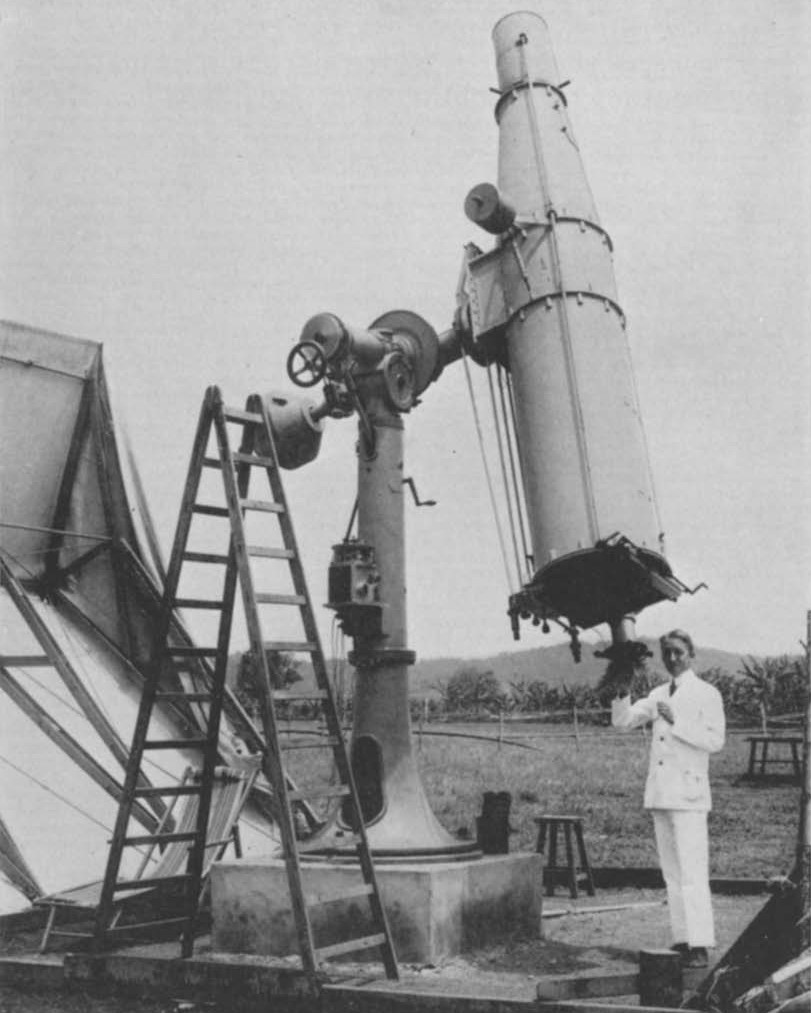}
\caption{The astrograph used in 1929 by the Potsdam observers. (Photo from \cite{Kluber1960})}
\end{subfigure}
\hfill
\begin{subfigure}[h]{0.45\linewidth}
\includegraphics[width=\linewidth]{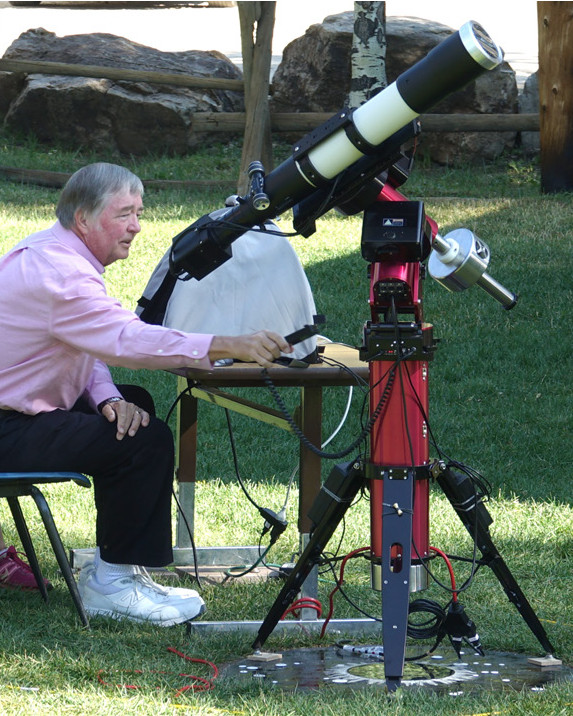}
\caption{The equipment used by D. Bruns in 2017. (Photo from \cite{Bruns2018})}\label{fig:bruns}
\end{subfigure}%
\caption{Nowadays, Einstein's theory can be verified with high accuracy using cheaper and smaller equipment. It is indeed fascinating how astronomers successfully managed to move tons of fragile material across the world during the first half of the past century.}
\end{figure}

\subsection{Dyson et al., 1919}
The first observation of the effect of General Relativity was carried out by British astronomers during the eclipse of May 29, 1919 \cite{Dyson1920,Kennefick2009}. Two expeditions were organized: Arthur Eddington and Edwin Turner Cottingham went to the island of Pr\'{i}ncipe, while the second team - formed by Andrew Crommelin and Charles Rundle Davidson - camped in Sobral, Brazil. Both expeditions observed stars belonging to the Hyades cluster in the constellation of Taurus, a region of the sky containing relatively bright stars.

On the day of the eclipse, Eddington took 16 plates but he had to face a sky full of clouds, which significantly reduced the quality of the images: when developed, 9 plates showed no signs of stars, and 4 plates had only faint, diffused marks. Luckily, clouds thinned for a while during the eclipse an two plates were usable. The analysis of the only five visible stars provided a deflection of $1.61 \pm 0.30 \textrm{ arcsec}$.

On the side of the Ocean, Sobral was favoured by fine weather and the team captured 19 plates with its main instrument, an astrographic telescope, plus 8 pictures using the backup instrument, a smaller 4-inch lens. However, the images obtained with the main instrument were diffused and apparently out of focus: hence, the authors decided to discard them, although they yielded to $0.93 \textrm{ arcsec}$, a deflection very close to the Newtonian (and not Einstein's) prediction. The second instrument, the 4-inch lens, performed well: its narrower field of view provided less stars on its plates, but the quality of images was higher and gave a deflection value of $1.98 \pm 0.16 \textrm{ arcsec}$.

\subsubsection{Chant and Young, 1922}
The Canadian astronomers Clarence Augustus Chant and Reynold Kenneth Young led an expedition to Wallal, western Australia, to observe the solar eclipse of September 21, 1922 \cite{Chant1923}.
The Canadian party captured two plates during the eclipse, while the night comparison plates were kindly captured by Campbell and Trumpler, members of the so-called Lick expedition.
On each eclipse plate, 18 stars were used to compute the light displacement: the obtained mean values ranged from $1.42 \textrm{ arcsec}$ to $2.16 \textrm{ arcsec}$, depending on the actual stars used for computation. Although the results agreed with the value predicted by Einstein, the accuracy of measurements was not sufficient to be considered decisive.

\subsubsection{Freundlich et al., 1929}
In 1929 the Potsdam observatory organized an expedition to Takengon, North Sumatra (now Indonesia) \cite{Freundlich1931}.
During the total solar eclipse of May 9, four images of the solar environment as well as three control plates were obtained. The astronomers noticed a very clear deflection of the light near the Sun. However, considering the displacement of 18 stars, they obtained an average value of $2.24 \pm 0.10 \textrm{ arcsec}$ at the Sun edge, which is noticeably larger than the theoretical value.

\subsection{Matukuma, 1939}
The total solar eclipse of June 19, 1936, gave Japanese astronomers an excellent opportunity for General Relativity verification. The expedition of the Tohoku Imperial University, led by professor Matukuma of the Astronomical Institute, placed their observations camp at Kosimizu, Abasiri \cite{Matukuma1941}.
Totality lasted less than two minutes, and during that time only one plate was taken. Ten stars within the Taurus constellation were recognized on the plate, but two of them were near the edges and could not be measured with enough accuracy. About six months later, two comparison plates (called No. 115 and 119) were taken at Sendai.
The average deflection value was $ 1,71 \textrm{ arcsec}$. However, measurement from plate 119 were quite unreliable: the author suggested an error while developing, drying or measuring the control plate.

\subsection{Van Biesbroeck, 1947}
In 1947 the Belgian-American astronomer George Van Biesbroeck took part in an expedition to Bocajuva, Brazil, to observe the total solar eclipse of May 20 \cite{Biesbroeck1950}. 
At the time of the eclipse, the Sun was located in front of one of the extended dark regions in Taurus. This lead to an unfavourable distribution of the stars: none of them was within less than 2.3 solar radii from the Sun's edge. Moreover, the auxiliary field stars were distorted due to the heating of the plane-parallel semi-transparent plate in front of the objective.
Two plates were taken in May and two check plates were acquired in August: 51 stars were considered, leading to a mean Einstein of $2.01 \pm 0.27 \textrm{ arcsec}$.

\subsection{Van Biesbroeck, 1952}
In 1952 George Van Biesbroeck travelled to Kartoum, Sudan, to make a new test \cite{Biesbroeck1953} with the same equipment previously used in Brazil in 1947.
On February 25, 1952 the astronomer exposed two plates. Differently from the first attempt, this time the plane-parallel plate was kept well shielded from the Sun and ventilated by an electric fan to insure equal distribution of temperature. Unfortunately, images were affected by the vibrations of the telescope due to a ``\emph{gusty wind}'' and many stars that were faintly visible had to be omitted. In total, a first plate showed 9 well measurable stars in the eclipse field, while a second had 11 measurable stars around the equatorial constellation of Aquarius.
Analysing the data, the astronomer obtained a relativity constant equal to $1.70 \pm 0.10 \textrm{ arcsec}$.

\subsection{Jones et al., 1973}
The Department of Astronomy and the Department of Physics of the University of Texas at Austin, in collaboration with the Department of Physics of Princeton University, sent an expedition to Chinguetti Oasis, Islamic Republic of Mauritania, to observe the Einstein shift at the total solar eclipse of June 30, 1973 \cite{Texas1976,Jones1976}. The eclipse lasted around six minutes and took place in a rich Milky Way field near the Gemini constellation. Three plates were obtained during the eclipse, while three sets of night-time plates of the same field were acquired five months later with the identical equipment.

All the plates were given to Burton F. Jones at the Royal Greenwich Observatory, who carried out careful plate reductions. During his analysis, Jones assigned to faint stars a much lower weight than the bright stars, computing weight as a function of magnitude. Eventually, 39 stars were considered and the computed light deflection value was $1.66 \pm 0.18 \textrm{ arcsec}$

\subsubsection{Bruns, 2017}
In 2017, Donald G. Bruns measured stars' deflection during the the August 21, 2017 total solar eclipse using high-quality amateur astronomical equipment \cite{Bruns2018}. A portable refractor, a CCD camera, and a computerized mount were set up in Casper, Wyoming, USA. 
A total of 45 images of the sky surrounding the Sun were acquired during totality. Two different techniques were used to analyse the data and two calibration fields were used to determine the plate scales.
A total of 20 stars were considered for measuring the deflection value, and the final result was $1.751 \pm 0.060 \textrm{ arcsec}$ - a value in perfect agreement with General Relativity and with the smallest uncertainty ever reported for this kind of experiment.

Bruns acknowledges that one of the key improvements that simplified this experiment was the availability of accurate absolute star positions. Hence, this experiment had less obstacles; nonetheless, it showed that nowadays Einstein's theory can be a pleasant learning experience without facing the costs, the risks and the difficulties encountered in the past by large expeditions.

\section{Raw and Processed Data}
As mentioned above, the aim of our work is the creation of a single, homogeneous dataset of all measurements acquired since 1919 during solar eclipses.
The outcome of our efforts is a 170-lines table, structured as shown in Table \ref{tab:dataset}.

\begin{table}[h]
\centering
\begin{tabular}{l c c c}
\hline
\textbf{Author (Year)} & \textbf{Star ID} & \textbf{Distance} & \textbf{Deflection} \\
\hline
2017 (Bruns) & TYC 832-663-1 & 3.395 & 0.508\\
2017 (Bruns) & TYC 836-444-1 & 2.433 & 0.757\\
2017 (Bruns) & TYC 836-948-1 & 3.555 & 0.353\\
2017 (Bruns) & TYC 832-142-1 & 1.513 & 1.312\\
2017 (Bruns) & TYC 833-1307-1 & 1.603 & 1.094\\
1973 (Jones) & SAO 78568 & 1.90 & 0.960\\
1973 (Jones) & SAO 78610 & 2.40 & 0.370\\
1973 (Jones) & SAO 78604 & 3.70 & 0.820\\
1973 (Jones) & ... & ...& \\
\hline
\end{tabular}
\caption{Excerpt of our dataset: distance and measured deflection are provided for each star.}\label{tab:dataset}
\end{table}

The first column identifies the year and the main observer who provided the data (see Section \ref{sec:expeditions} for more information on the expeditions).
The second column contains the ID of the star: for older catalogues, we converted the references to a more recent catalogue (SAO, Tycho, or Bonner Durchmusterung) and we wrote the identifier in a string compatible both with the SIMBAD Astronomical Database \cite{simbad} and the educational program Stellarium \cite{stellarium}.
The last two columns contain the distance of the star from Sun - expressed in solar radii - and the measured deflection, expressed in \textit{arcsec}.

This table contains only one value for each star: when multiple measurements where available for one star, we averaged them to obtain a single value. Combined with this homogeneous database of measurements, we provide all the original values as separate files. Hence, the raw value are still available for further analysis.

We decided to make all data freely available online using the collaborative platform GitHub - this platform is widespread, reliable, and offers a simple way to signal issue and suggest edits. The whole dataset of measurements is available for download at \cite{github}.

\begin{figure}[!htb]
\centering\includegraphics[width=0.7\linewidth]{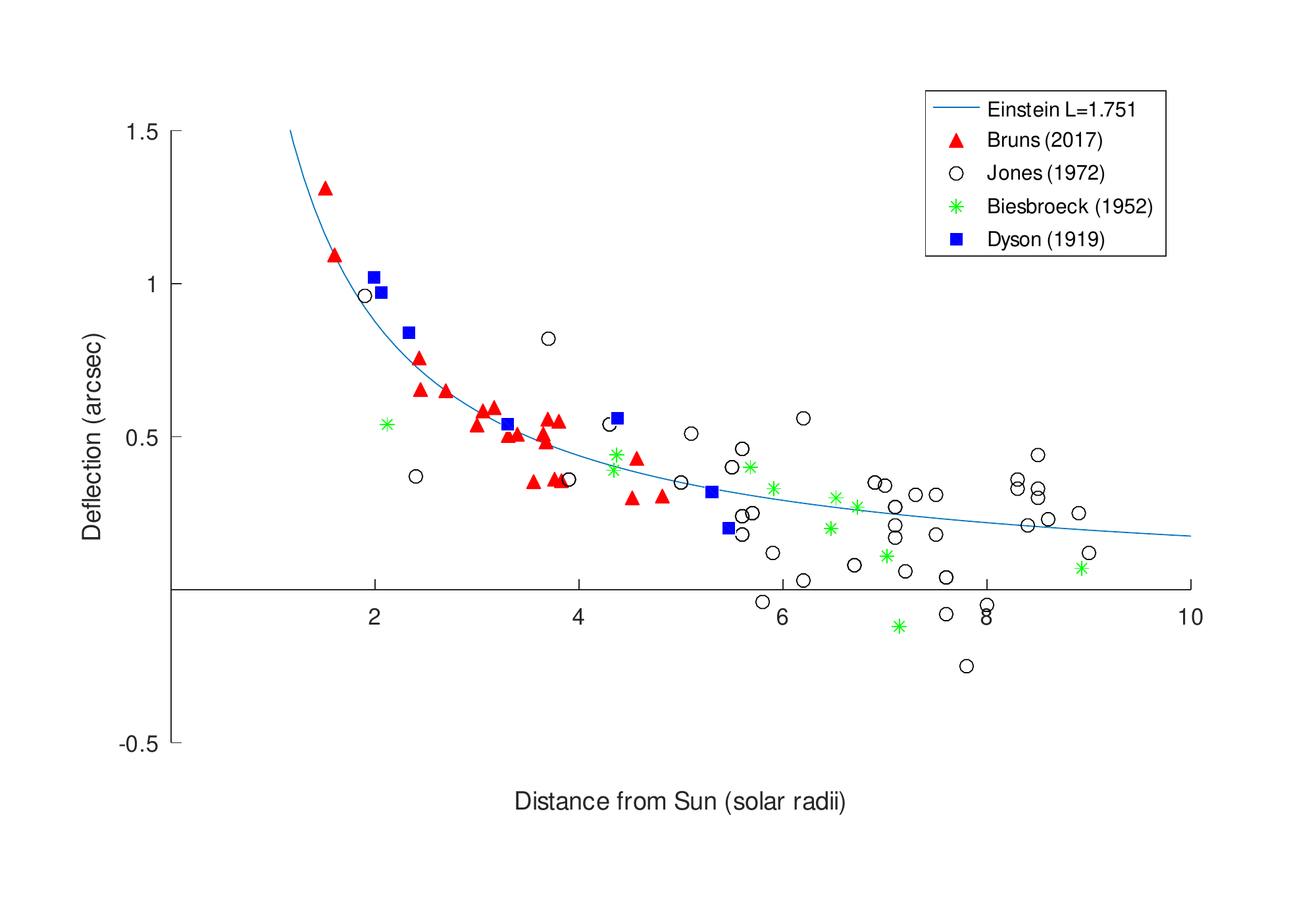}
\caption{Sample plot combining measurements taken in 1919, 1952, 1972, and 2017 - comparing the results obtained during the last century is a matter of minutes when they are all available in single database.}\label{fig:plot}
\end{figure}

The dataset is provided in CSV format, so that is can be easily imported by most of spreadsheet programs, data analysis tools, and programming languages (Figure \ref{fig:plot}). In addition, a folder for each expedition contains the raw values in CSV format, a brief text document describing how the data has been processed when adding them to the main dataset, plus a link to the original article.
When additional scripts have been used to process the measurements, we published the source code within the appropriate folder.

\section{Classroom Activity}
Here we propose a possible group activity for high-school students. 
We believe it is very important to be as much general (and anonymous) as possible when presenting the task, in order to keep students unbiased toward the data. We also suggest to create a playful learning environment, starting the lesson more or less as follows:
\textit{``Two famous scientists A and B have proposed two alternative theories for an important astronomical phenomenon.
According to A, the actual value should be $1.75"$; on the contrary, B's theory leads to $0.87"$. The scientific community is split: different teams around the world have been asked to carry out experiments and gather useful data. 
Today we have received all the values obtained on the field. Now we have been asked to analyse them and choose which theory is the right one (or if both are wrong). The world is waiting for us!''}

Then, each group of students will receive a portion the data coming from an expedition, and they will let be free to analyse the data using the methods and the tools they are more comfortable with (such as graphical interpolation, spreadsheet, ad-hoc computer programs, etc.). Eventually, after a few days, they will be asked to present their conclusions to the classroom.

We believe that the combination of historical, numerical and team-work factors represents an opportunity both to present General Relativity theory in physics classes in an engaging way and to show how scientists tried during the last century to verify (or disprove) a fundamental theory.

\section{Conclusion}
More than one hundred years since its first observation, the idea that Sun's gravity makes ``\emph{lights all askew in the heavens}'' is still fascinating.
To celebrate the centenary of the first observation of gravitational bending for starlight that grazes the Sun, we collected dozens of measurements taken from 1919 to 2017 in single dataset of homogeneous values. 
Together with the whole dataset, now freely available on-line, we suggested a playful activity for high-school students. We  would like to extended this work including more activities and exercises: hence, we invite the reader to use the data with young scientists and to provide us feedback and suggestions.


\section*{Acknowledgements}
We would like to thank Davide Cavalca and Gabriele Marangoni for giving the paper a critical reading and for providing several helpful comments. We are also grateful to the Accademia Nazionale Virgiliana and to the Mantova chapter of AIF (Associazione per l'Insegnamento della Fisica, the Italian association for physics teaching) for their support and for giving us the opportunity to meet several teachers and to exchange ideas with them.

\bibliographystyle{model1-num-names}
\bibliography{eclipses.bib}







\end{document}